\documentclass[lettersize,journal]{IEEEtran}
\usepackage{amsmath}
\allowdisplaybreaks
\usepackage{amssymb}
\usepackage{amsfonts}
\usepackage{algorithmicx}
\usepackage{algorithm}

\usepackage{algpseudocode}
\usepackage{array}
\usepackage[caption=false,font=normalsize,labelfont=sf,textfont=sf]{subfig}
\usepackage{textcomp}
\usepackage{stfloats}
\usepackage{url}
\usepackage{verbatim}
\usepackage{subcaption}
\usepackage{graphicx}
\usepackage{cite}
\usepackage{tikz}
\usetikzlibrary{shapes,arrows,positioning,calc,decorations.pathreplacing}
\begin{document}

\title{Hardware-Impaired Over-the-Air Computation with Fluid Antenna Array}

\author{{Zilong Li, Jianxin Dai,~\IEEEmembership{Senior Member, IEEE}, Zhaohui Yang,~\IEEEmembership{Member, IEEE}, Zhaoyang Zhang, 

Kai-Kit Wong,~\IEEEmembership{Fellow, IEEE}, Zhiyang Li, and Shunkuan Cheng }
\thanks{Jianxin Dai is with the School of Science, Nanjing University of Posts
and Telecommunications, Nanjing 210023, China (e-mail: daijx@njupt.edu.cn).}
\thanks{Zilong Li is with the School of Science, Nanjing University of Posts and Telecommunications, Nanjing 210023, China (email: 1024
082207@njupt.edu.cn).}
\thanks{Zhaohui Yang, and Zhaoyang Zhang are with the College of Information Science and Electronic Engineering, Zhejiang University, Zhejiang 310027, China, and also with Zhejiang Provincial
Key Laboratory of Information Processing, Communication Networking (IPCAN), Hangzhou 310027, China (e-mail: yang$\_$zhaohui@zju.edu.cn; ning$\_$ming@zju.edu.cn).}
\thanks{Kai-Kit Wong is with the Department of Electronic and Electrical
 Engineering, University College London, WC1E 6BT London, U.K., and also
 with the Yonsei Frontier Laboratory, Yonsei University, Seoul 03722, South
 Korea (e-mail: kai-kit.wong@ucl.ac.uk).}
\thanks{Zhiyang Li and Shunkuan Cheng are with China Mobile (Hangzhou) Information Technology Company Ltd., Hangzhou 311121, China (e-mails: lizhiyang@cmhi.chinamobile.com; chengshunkuan@cmhi.chinamobile.com).}}

\maketitle
\begin{abstract}
 This paper investigates a fluid antenna (FA) array-enhanced over-the-air computation (AirComp) system in the presence of hardware impairments (HWIs), exploiting the new degrees of freedom offered by reconfigurable antenna positioning. To minimize the mean squared error (MSE) of the aggregated signal, we jointly optimize  transmit power control, receive beamforming, and the antenna position vector (APV), subject to practical constraints such as HWI-induced distortion noise, FA movement energy consumption, and total power budgets. The resulting optimization problem is non-convex and highly coupled. To address it efficiently, we adopt a block coordinate descent (BCD) framework, decomposing it into three manageable subproblems.  For each subproblem, closed-form solutions or efficient numerical algorithms are derived. Simulation results demonstrate that the proposed joint transceiver and APV design significantly reduces the MSE compared to conventional fixed-position antenna (FPA) arrays and exhibits enhanced robustness against hardware impairments. The effectiveness and convergence of the proposed algorithm are further validated under various system configurations.
\end{abstract}

\section{Introduction}
\IEEEPARstart{T}{he} proliferation of intelligent services, driven by rapid progress in artificial intelligence, is creating an unprecedented need for large-scale connectivity and efficient data aggregation \cite{ref1}. The fulfillment of these requirements is,  however, significantly challenged by limited radio resources and strict latency constraints. In this context, over-the-air computation (AirComp) presents itself as a transformative technology\cite{ref2,ref3}. AirComp fundamentally leverages the inherent superposition property of wireless multiple-access channels to merge communication and computation processes. This principle enables a revolutionary multiple-access paradigm shift from “compute-after-communicate” to “compute-when-communicate\cite{ref4}.”

Prior studies have extensively explored AirComp across various application scenarios. Specifically, the authors in \cite{ref5} and \cite{ref6} designed and optimized schemes based on AirComp technology, which were respectively applied to the scenarios of edge-device co-inference and communication-efficient federated edge learning, with the core goal of enhancing communication efficiency and the performance of related tasks. Subsequent research by \cite{ref7} and \cite{ref8} based on AirComp technology, targeted wireless data aggregation, optimized performance for Internet of things (IoT) scenarios—the former focused on massive IoT while the latter on mobile sensors—with the core of integrating communication and computation to improve data aggregation efficiency and reduce computation errors. 

Fluid antenna (FA) and movable antenna (MA) systems represented a significant technological advancement capable of creating new degrees of freedom (DoFs) in wireless systems \cite{ref9,ref10,ref11}. By allowing precise antenna repositioning, these systems enabled dynamic optimization of channel conditions, presenting a new solution. The authors in \cite{ref12}, by integrating cutting-edge fluid antenna hardware with efficient Aircomp algorithms, paved a promising new pathway for the efficient and accurate processing of massive data in future wireless networks. Building on this foundation, the first study \cite{ref13} pioneered robust resource optimization for FA systems in AirComp under practical channel uncertainty constraints, significantly enhancing interference resistance in non-ideal environments.

Against the above backdrop, this paper aims to address the challenge of accurate AirComp in practical non-ideal hardware environments by leveraging the flexibility of FA arrays. Our primary objective is to minimize the MSE of the aggregated signal. This is achieved by formulating and solving a joint optimization problem that encompasses the transceiver design (i.e., transmit power allocation and receive beamforming) and the FA position vector, all while accounting for critical practical constraints including hardware distortion noise, the energy cost of moving antennas, and power budgets.

\section{System Model and Problem Formulation}
\begin{figure}[t]
    \centering
    \begin{minipage}[b]{0.24\textwidth}
        \centering
        \includegraphics[width=\textwidth]{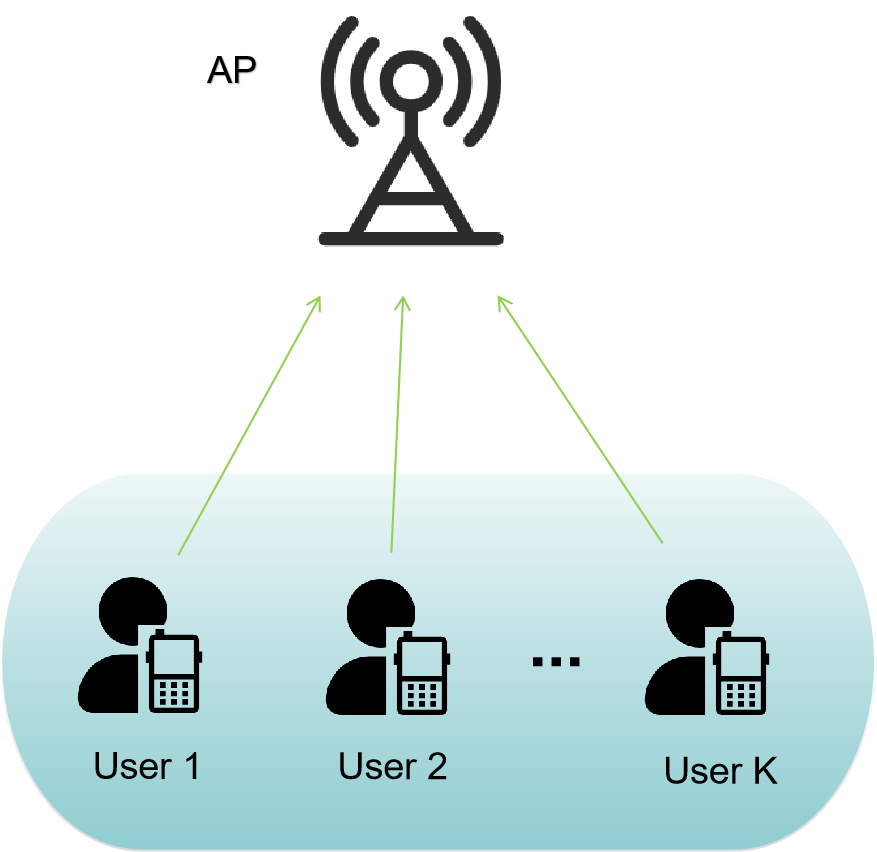}
        \par\vspace{2pt} % 添加一点垂直间距
        \centering (a) System model
        \label{a:a}
    \end{minipage}
    \hfill
    \begin{minipage}[b]{0.24\textwidth}
        \centering
        \includegraphics[width=\textwidth]{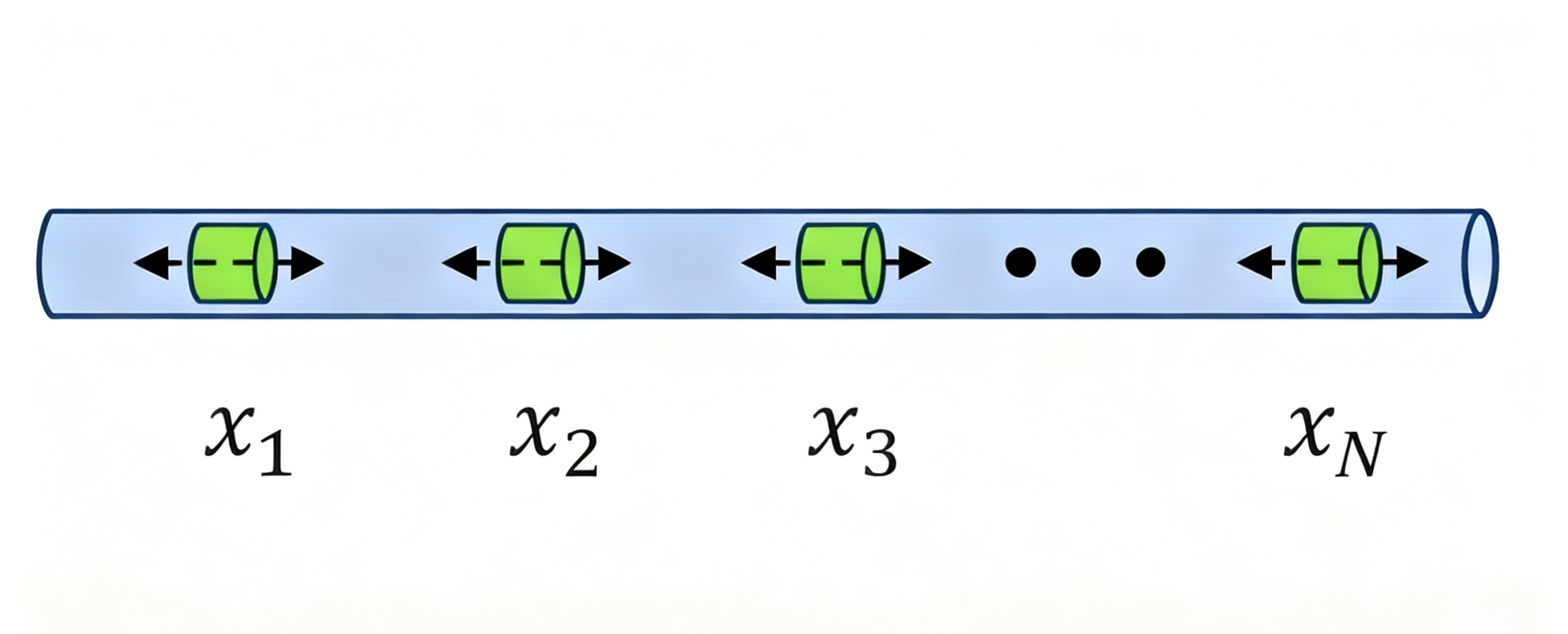}
        \par\vspace{2pt} % 添加一点垂直间距
        \centering (b) FA array
        \label{fig:image2}
    \end{minipage}
    \caption{AirComp system enhanced by FA.}
    \label{fig:both}
\end{figure}
\subsection{System Model}
An uplink multi-user single-input multiple-output (SIMO) communication system is considered, where an FA array equipped
with $N$ FAs is deployed to assist in uplink communications from $K$ single-antenna users to an access point (AP) with $N$ antennas. Let $s_k {\in}{\mathbb{C}}$ represent the data generated by user $k$, $\forall k  {\in}       \mathcal{K}{\triangleq}\{1,...,K\}$. For simplicity, we assume that $s_k$  possesses zero mean and unit power, i.e., $E[s_k] = 0, E[s_ks^H_k]
 = 1, \forall k \in \mathcal{K}$, and $E[s_ks^H_i]=0,\forall k\neq i$.
The sum of all users' data 
\begin{equation}
s=\sum_{k=1}^K{s_k}.
\end{equation}
is computed at the AP by harnessing the superposition property of the
wireless multiple-access channel.

The FAs’ positions can be adjusted in the given one-dimensional (1$\text{D}$) line segment of length $L$. Let $x_n \in [0, L]$ denote the $n$-th FA’s position, and the APV of all $N$ FAs is denoted by $\boldsymbol{x} \triangleq [x_1, x_2, . . . , x_N ]^T\in \mathbb{R}^{N\times1}$ with $0 \leq x_1 < x_2 <...< x_N \leq L$ without loss of generality. Thus, the steering vector of the MA array can be written as a
function of the APV $\boldsymbol{x}$ and the steering angle $\theta$:
\begin{equation}
    \boldsymbol{a}(\boldsymbol{x},\theta)=[e^{j\frac{2\pi}{\lambda}x_1\cos(\theta)},...,e^{j\frac{2\pi}{\lambda}x_N\cos(\theta)}]^T\in \mathbb{C}^{N\times 1},
\end{equation}
where $\lambda$ denotes the wavelength.

Assuming that the position adjustment of FAs is realized by mechanical devices, the movement energy consumption is proportional to the movement distance, which can be defined as:
\begin{equation}
    P_{\text{move}}=\sum_{n=1}^N\xi|x_n-x^{\text{init}}_n|,
\end{equation}
where $\xi$ is the energy consumption coefficient per unit distance, $x^{\text{init}}_n$ is the initial position of the $\textit{n}$-th FA, and $|x_n-x^{\text{init}}_n|$ is its movement distance. 

Let $\boldsymbol{h}_k \in \mathbb{C}^{N\times1}$ denote the channel from
user $\textit{k}$ to AP, given by
\begin{equation}
    \boldsymbol{h}_k=\alpha_k\boldsymbol{a}(\boldsymbol{x},\theta_k), k\in \mathcal{K},
\end{equation}
where $\alpha_k$ and $\theta_k$ represent the propagation gain and angle of arrival (AoA) of the line-of-sight (LOS) path of user $k$, respectively. 

The received signal at the AP considering the realistic hardware impairments is then expressed as
\begin{equation}
   \boldsymbol{y}=\underbrace{\sum_{k=1}^K{\boldsymbol{h}_kw_ks_k+\boldsymbol{n}}}_{\boldsymbol{\tilde{y}}}+\boldsymbol{\gamma},
\end{equation}
where $\boldsymbol{\tilde{y}} \in \mathbb{C}^{N\times 1}$ denotes the undistorted received signal, $w_k$ is the transmit equalization coefficient of user $k$, $\forall k \in \mathcal{K}$, and $\boldsymbol{n} \sim \mathcal{CN}(0,\sigma^2\boldsymbol{I})$ represents the additive white Gaussian noise (AWGN) at the AP. $\boldsymbol{\gamma}\sim \mathcal{CN}(0,\beta^2\widetilde{\text{diag}}(E[\boldsymbol{\tilde{y}}\boldsymbol{\tilde{y}}^H])
$ denotes the AP distortion noise, where $\beta$ indicates the distortion level at the AP and $\widetilde{\text{diag}}(E[\boldsymbol{\tilde{y}}\boldsymbol{\tilde{y}}^H])$ denotes a diagonal matrix formed by extracting the main diagonal elements of matrix $E[\boldsymbol{\tilde{y}}\boldsymbol{\tilde{y}}^H]$.

The estimated target function, equipped with a receive beamforming vector $\boldsymbol{m} \in \mathbb{C}^{N\times1}$ at the AP, can be formulated as
\begin{equation}
\tilde{s}=\boldsymbol{m}^H\boldsymbol{y}=\sum_{k=1}^K{\boldsymbol{m}^H\boldsymbol{h}_kw_ks_k}+\boldsymbol{m}^H\boldsymbol{n}+\boldsymbol{m}^H\boldsymbol{\gamma} .
\end{equation}
\subsection{Problem Formulation}
In this section, we examine the MSE assuming the perfect CSI when the receiver is under the effect of HWIs. The MSE for minimizing the error between the target and estimated function variables can be obtained as:
\begin{multline}
\text{MSE}=E[|\tilde{s}-s|^2]
=\sum_{k=1}^K{|\boldsymbol{m}^H\boldsymbol{h}_kw_k-1|^2}\\
+\sigma^2||\boldsymbol{m}||^2+\beta^2\boldsymbol{m}^H\widetilde{\text{diag}}(\sum_{k=1}^K{|w_k|^2\boldsymbol{h}_k\boldsymbol{h}_k^H+\sigma^2\boldsymbol{I})\boldsymbol{m}}.
\end{multline}

Our goal is to jointly optimize the transmit equalization coefficient of all users $\boldsymbol{w}\triangleq [w_1,..., w_K]^T$, the receive beamforming vector $\boldsymbol{m}$ at the AP, and the location vector $\boldsymbol{x}$ of the FA array to minimize MSE in (7). The specific optimization problem for computing MSE is formulated as follows.
\begin{subequations}\label{eq:2}
\begin{align}
\min\limits_{\boldsymbol{w},\boldsymbol{m},\boldsymbol{x}}\quad&\text{MSE} \label{eq:2A}\\
s.t.\quad &|w_k|^2\leq P_k, \forall  k \in \mathcal{K} ,\label{eq:2B}\\
&P_{\text{move}}+\sum_{k=1}^K{|w_k|^2}\leq P_{\text{total}}, \label{eq:2c}\\
&x_1\geq0,  x_N\leq L, \label{eq:2d}\\
&x_n-x_{n-1}\geq L_0,\forall n=2,...,N.\label{eq:2e}
\end{align}
\end{subequations}
where (8b) ensures that each user’s transmission
power does not exceed its maximum allowable power and (8c) is the total energy constraint combining mobile energy consumption and user transmission energy consumption, of which $P_{\text{total}}$ is the total allowable energy upper limit for the system, the constraint (8d) guarantees that the FAs are moved within the feasible region $[0,L]$. $L_0$ in constraint (8e) is the minimum distance between adjacent FAs to avoid antenna coupling.
\begin{algorithm}[t]
\caption{ Algorithm for Transmit Power Allocation Optimization}
\begin{algorithmic}[1]
\State \textbf{Initialize:} $\lambda_L = 0$, $\lambda_U = 1$, iteration number $n = 0$, $den_k = |a_k|^2+\beta^2c_k$.
\State Compute initial $w_k^{(0)} = \frac{\bar{a}_k}{den_k}$ for all $k \in \mathcal{K}$.
\Repeat
\State Set $n = n + 1$
\For{each $k \in \mathcal{K}$ with $|w_k^{(n-1)}|^2 > P_k$}
\State Update $w_k^{(n)} = \frac{\sqrt{P_k}}{|\bar{a}_k|}\bar{a}_k$
\EndFor
\While{$\sum_{k=1}^K \left|\frac{\bar{a}_k}{den_k + \lambda_U}\right|^2 > P_{\text{budget}}$}
\State Update $\lambda_U = 2\lambda_U + 1$
\EndWhile
\For{$iter = 1$ to $60$}
\State Update $\lambda = 0.5(\lambda_L + \lambda_U)$
\State Update $w_k^{(n)} = \frac{\bar{a}_k}{den_k + \lambda}$ for all $k$
\If{$\sum_{k=1}^K |w_k^{(n)}|^2 \leq P_{\text{budget}}$}
\State Update $\lambda_U = \lambda$
\Else
\State Update $\lambda_L = \lambda$
\EndIf
\EndFor
\Until{$\sum_{k=1}^K |w_k^{(n)}|^2 \leq P_{\text{budget}}$ and $|w_k^{(n)}|^2 \leq P_k$ for all $k$}.
\end{algorithmic}
\end{algorithm}

\section{Transceiver and APV Design}
In this section, we utilize the concept of block coordinate
descent to tackle the coupling between $\boldsymbol{w}$, $\boldsymbol{m}$ and $\boldsymbol{x}$ in (8), which involves sequentially optimizing certain variables while holding the remaining variables constant.
\subsection{Subproblem 1: Transmit Power Allocation}
Following this, we optimize $\boldsymbol{w}$ under given other parameters. The
associated optimization problem with respect to  $\boldsymbol{w}$ can be
formulated as follows:
\begin{align}
\min\limits_{\boldsymbol{w}}\quad&\sum_{k=1}^K({|\boldsymbol{m}^H\boldsymbol{h}_kw_k-1|^2}+{|w_k|^2\beta^2\boldsymbol{m}^H\widetilde{\text{diag}}(\boldsymbol{h}_k\boldsymbol{h}_k^H)\boldsymbol{m}}) \nonumber \\
\text{s.t.}\quad &(8b),(8c).
\end{align}

Since the objective function and the constraints are all convex, Problem (9) is a convex problem. We introduce Lagrangian multipliers $\lambda \geq 0$ to construct the Lagrangian function:
\begin{align}
 \mathnormal{\mathcal{L}}(\boldsymbol{w}, \lambda) = &\sum_{k=1}^{K} \left( |a_k w_k - 1|^2 + \beta^2 c_k |w_k|^2 \right) + \nonumber \\
 &\lambda\left(\sum_{k=1}^{K} |w_k|^2 - P_{\text{budget}}\right),
\end{align}
where  $a_k=\boldsymbol{m}^H \boldsymbol{h}_k$, $c_k = \boldsymbol{m}^H \widetilde{\text{diag}}(\boldsymbol{h}_k{\boldsymbol{h}_k}^H)\boldsymbol{m}$, and $P_{\text{budget}} = P_{\text{total}} - P_{\text{move}}$. 

For the convex optimization problem (9), the point $(\boldsymbol{w}^*, \lambda^*)$ is optimal if and only if:
\begin{subequations}\label{eq:2}
\begin{align}
&\nabla_{\boldsymbol{w}}\mathnormal{{\cal L}}({\boldsymbol{w}}^{*},\lambda^{*})=0,  \label{eq:2A}\\
&|w_{k}^{*}|^{2}\leq P_{k},\ \sum_{k=1}^{K}|w_{k}^{*}|^{2} \leq P_{\rm budget},\label{eq:2B}\\
&\lambda^{*}\left(\sum_{k=1}^{K}|w_{k}^{*}|^{2}-P_{\rm budget}\right)=0,  \label{eq:2c}\\
&\lambda^{*}\geq 0, \label{eq:2d}
\end{align}
\end{subequations}
where $\mathnormal{\mathcal{L}}$ is the Lagrangian function defined in (10).

From the stationarity condition (11a), we have:
\begin{align}
\frac{\partial \mathit{\mathcal{L}}}{\partial {w}_k} &= \bar{a}_k(a_k w_k^* - 1) + (\beta^2 c_k + \lambda^*)w_k^* = 0 .
\end{align}
Then the optimal transmit power coefficients $\boldsymbol{w}^*$ satisfy the following closed-form expression:
\begin{align}
    w_k^*(\lambda^*) = \frac{\bar{a}_k}{|a_k|^2 + \beta^2 c_k + \lambda^*}, \quad \forall k \in \mathcal{K},
    \label{eq:optimal_solution}
\end{align}
where $\lambda^*$ is the optimal Lagrange multiplier satisfying the complementary slackness conditions.

To determine the power allocation for beamforming vectors under total power budget and per-user power constraints, we propose an iterative algorithm based on bisection search. This algorithm progressively converges to a feasible solution satisfying all constraints by adjusting the Lagrangian multiplier. The specific Algorithm 1 is as follows.

\subsection{Subproblem 2: Receiver Beamforming Design}
We proceed to optimize $\boldsymbol{m}$ in (8) with the $\boldsymbol{b}$ and $\boldsymbol{x}$ fixed. This involves solving the following unconstrained convex optimization problem:
\begin{equation}
\min\limits_{\boldsymbol{m}}\text{MSE}.
\end{equation}

It is observed that (17) is a (convex) least squares problem, we compute the gradient of the objective function with respect to $\boldsymbol{m}$ and set it to zero, the optimal solution for the $\boldsymbol{m}$ is given by
\begin{equation}
\boldsymbol{m}^*=\boldsymbol{R}^{-1}\sum_{k=1}^K{w_k\boldsymbol{h}_k},
\end{equation}
where $\boldsymbol{R}=(1+\beta^2)\sigma^2\boldsymbol{I}+\beta^2\widetilde{\text{diag}}(\sum_{k=1}^K{|w_k|^2\boldsymbol{h}_k\boldsymbol{h}_k^H})+\sum_{k=1}^K{|w_k|^2\boldsymbol{h}_k\boldsymbol{h}_k^H}$.

\subsection{Subproblem 3: FA Array Design}
The associated optimization problem with respect to $\boldsymbol{x}$ is
given by
\begin{align}
\min\limits_{\boldsymbol{x}}\quad&\text{MSE} \nonumber\\
\text{s.t.}\quad &(8c),(8d),(8e).
\end{align}

This objective function is highly non-convex. To address it,
we employ a projected gradient descent method, integrated with an Armijo line search condition and a budget constraint handling mechanism.

Due to the complex analytical form of the objective function involving channel response calculations and mean square error evaluations, the gradient is computed numerically using the forward finite difference method. For each coordinate direction 
$n$, the partial derivative is approximated as:
\begin{align}
g_n &=  \frac{\text{MSE}(\boldsymbol{x} + \epsilon \boldsymbol{e}_n) - \text{MSE}(\boldsymbol{x})}{\epsilon}=\frac{1}{\epsilon} \Bigg[ \nonumber \\
 &  \sum_{k=1}^{K} \left( \left| \boldsymbol{m}^{H} \boldsymbol{h}_{k}(\boldsymbol{x} + \epsilon \boldsymbol{e}_{n}) w_{k} - 1 \right|^{2} - \left| \boldsymbol{m}^{H} \boldsymbol{h}_{k}(\boldsymbol{x}) w_{k} - 1 \right|^{2} \right) \nonumber \\
& + \beta^{2} \boldsymbol{m}^{H} \left( \widetilde{\text{diag}} \left( \sum_{k=1}^{K} |w_{k}|^{2} \boldsymbol{h}_{k}(\boldsymbol{x} + \epsilon \boldsymbol{e}_{n}) \boldsymbol{h}_{k}^{H}(\boldsymbol{x} + \epsilon \boldsymbol{e}_{n}) \right) \right. \nonumber \\
& \left. - \widetilde{\text{diag}} \left( \sum_{k=1}^{K} |w_{k}|^{2} \boldsymbol{h}_{k}(\boldsymbol{x}) \boldsymbol{h}_{k}^{H}(\boldsymbol{x}) \right) \right) \boldsymbol{m} \Bigg],
\end{align}
 where $\epsilon = 10^{-4}\lambda$ represents a carefully chosen perturbation size scaled by the wavelength parameter $\lambda$, and $e_n$ denotes the standard basis vector. This approach provides a general-purpose gradient approximation that adapts to the problem's inherent scale while maintaining numerical stability.We use the  projected gradient descent method to obtain a locally optimal $\boldsymbol{x}$.
\begin{algorithm}[ht]
\caption{Algorithm for APV Positions Optimization}
\begin{algorithmic}[1]
\State \textbf{Initialize:} $x_n^{(0)} = \frac{L(n-1)}{N}$, $\alpha^{(0)} = 0.7\lambda$, $k = 0$, $P_{\text{budget}_x} = P_{\text{total}}-P_{\text{move}}$
\State Compute initial channel matrix $\boldsymbol{H}^{(0)}$
\State Compute initial objective function $F^{(0)}$
\State Compute initial gradient $\boldsymbol{g}^{(0)}$ via finite differences

\Repeat
\State Set $k = k + 1$
\State Update $\boldsymbol{x}^{(k)} = \boldsymbol{x}^{(k-1)} - \alpha^{(k-1)} \cdot \boldsymbol{g}^{(k-1)}$

\State \textbf{Projection:}
\State Update $\boldsymbol{x}^{(k)} = \text{sort}(\boldsymbol{x}^{(k)})$
\State Update $\boldsymbol{x}^{(k)} = \max(\min(\boldsymbol{x}^{(k)}, L), 0)$
\For{$i = 2$ to $N$}
    \If{$\boldsymbol{x}_i^{(k)} < \boldsymbol{x}_{i-1}^{(k)} + d_{\min}$}
        \State Update $\boldsymbol{x}_i^{(k)} = \boldsymbol{x}_{i-1}^{(k)} + d_{\min}$
    \EndIf
\EndFor

\While{$\sum_{n=1}^N\xi|x_n^{(k)}-x_n^{(0)}| > P_{\text{budget}_x}$}
    \State Update $\alpha^{(k-1)} = 0.5 \cdot \alpha^{(k-1)}$
    \State Update $\boldsymbol{x}^{(k)} = \boldsymbol{x}^{(k-1)} - \alpha^{(k-1)} \cdot \boldsymbol{g}^{(k-1)}$
    \State \textbf{Projection:}
    \State Update $\boldsymbol{x}^{(k)} = \text{sort}(\boldsymbol{x}^{(k)})$
    \State Update $\boldsymbol{x}^{(k)} = \max(\min(\boldsymbol{x}^{(k)}, L), 0)$
    \For{$i = 2$ to $N$}
        \If{$\boldsymbol{x}_i^{(k)} < \boldsymbol{x}_{i-1}^{(k)} + d_{\min}$}
            \State Update $\boldsymbol{x}_i^{(k)} = \boldsymbol{x}_{i-1}^{(k)} + d_{\min}$
        \EndIf
    \EndFor
    \If{$\alpha^{(k-1)} < 10^{-6}$} \textbf{break} \EndIf
\EndWhile

\State Compute $\boldsymbol{H}^{(k)}$ and $F^{(k)}$

\If{$F^{(k)} \leq F^{(k-1)} - \eta \cdot \alpha^{(k-1)} \cdot \|\boldsymbol{g}^{(k-1)}\|^2$}
    \State Update gradient $\boldsymbol{g}^{(k)}$ via finite differences
    \State Update $\alpha^{(k)} = \min(1.2 \cdot \alpha^{(k-1)}, 0.7\lambda)$
\Else
    \State Update $\boldsymbol{x}^{(k)} = \boldsymbol{x}^{(k-1)}$
    and $F^{(k)} = F^{(k-1)}$
    \State Update $\alpha^{(k)} = 0.5 \cdot \alpha^{(k-1)}$
\EndIf
\Until{$k \geq 30$ or $\alpha^{(k)} < 10^{-6}$}
\end{algorithmic}
\end{algorithm}
Algorithm 2 outlines the approach for designing the proposed transceiver and APV. Its convergence is guaranteed since
in each iteration, the objective function declines or remains
unchanged.

The proposed transceiver and APV design approach is
outlined in Algorithm 3. Moreover, the convergence of
Algorithm 3 is illustrated in the following theorem.

Theorem 1: The proposed BCD algorithm 3 is guaranteed to converge to a stationary point of Problem (8).

Proof: Algorithm 3 sequentially optimizes three variable blocks. To prove the convergence of Algorithm 3, we introduce superscript $t$ as the iteration index, e.g., $\boldsymbol{m}^t$ represents the decoding vector at the end of the $t$-th iteration round. Then, Algorithm 3 converges as
\begin{align}
\text{MSE}(\boldsymbol{w}^t, \boldsymbol{m}^t, \boldsymbol{x}^t) &\overset{(a)}{\geq} \text{MSE}(\boldsymbol{w}^{t+1}, \boldsymbol{m}^t, \boldsymbol{x}^t) \nonumber \\
&\overset{(b)}{\geq} \text{MSE}(\boldsymbol{w}^{t+1}, \boldsymbol{m}^{t+1}, \boldsymbol{x}^t) \nonumber \\
&\overset{(c)}{\geq} \text{MSE}(\boldsymbol{\boldsymbol{w}}^{t+1}, \boldsymbol{m}^{t+1}, \boldsymbol{x}^{t+1}),
\end{align}
where (a), (b), and (c) follow since the updates of $\boldsymbol{w}$,
$\boldsymbol{m}$, and $\boldsymbol{x}$ are the optimal (or locally optimal) solutions
to (9), (17), and (19), respectively. The optimal solution for each subproblem ensures that the objective function value is monotonically non-increasing. Since the MSE is lower-bounded by zero, the algorithm is guaranteed to converge. 

Next, we analyze the computational complexity of the algorithm as follows. Let $I$, $I_{\boldsymbol{w}}$, $I_{\boldsymbol{m}}$, and $I_{\boldsymbol{x}}$ denote the maximum number of iterations for the outer BCD loop, the maximum number of backtracking iterations for receive beamforming optimization, the maximum number of iterations for bisection search in transmit power allocation, and the maximum number of iterations for gradient descent in FA position optimization, respectively. The computational complexity order for receive beamforming optimization is $\mathcal{O}(I_{\boldsymbol{m}} K N^2)$, for transmit power allocation is $\mathcal{O}(I_{\boldsymbol{w}} K)$, and for FA position optimization is $\mathcal{O}(I_{\boldsymbol{x}} K N^3)$. Therefore, the overall complexity order of the algorithm is $\mathcal{O}\big(I(I_{\boldsymbol{w}} K + I_{\boldsymbol{m}} K N^2 + I_{\boldsymbol{x}} K N^3)\big)$.
\begin{algorithm}[t]
\caption{Pseudo-code for the proposed design of
transceivers and APV}
\label{alg:proposed_design}
\begin{algorithmic}[1]
\State \textbf{Initialization:}
\State \quad Set initial receive beamforming vector: $\boldsymbol{m} = e^{j \angle(\cdot)}$
\State \quad Set initial APV positions: $x_n = \frac{L(n-1)}{N}$, $\forall n \in \{1, \ldots, N\}$
\While{not converged}
\State $i \leftarrow i + 1$
\State \textbf{Step 1: Update Power Allocation Coefficients}
\State \quad Update $\boldsymbol{w}_i$ by solving equation (9)
\State \textbf{Step 2: Update Receive Beamforming Vectors}
\State \quad Update $\boldsymbol{m}_i$ using equation (18)
\State \textbf{Step 3: Update APV Positions}
\State \quad Update $\boldsymbol{x}_i$ using the projected gradient descent method
\EndWhile
\end{algorithmic}
\end{algorithm}
\section{Simulation Results}
In this section, the numerical results are provided to evaluate
the performance of the proposed algorithm for an FAs-aided
AirComp system with hardware impairments. We set $L_0 = 0.5 \lambda$, $L = N \lambda$, $P_k = P_0$, $\forall k \in \mathcal{K}$, $\beta = 0.8$, and  $\xi = 0.8$. We compare the performance
of our proposed designs with two benchmark schemes: one
that ignores HWIs and another that employs FPA. For the
benchmark that ignores HWIs, we optimize the transceiver design by solving the optimization problem
without accounting for hardware impairments. For the FPA
benchmark, the positions of $N$ FAs are uniformly distributed
across the feasible region $[0, L]$. This setup fixes the APV
as $\boldsymbol{x} = [ 0, \frac{L}{N}, \ldots, \frac{(N-1)L}{N} ]^T$. The optimal configurations for $\boldsymbol{w}$ and $\boldsymbol{m}$ are determined iteratively using (9) and (18) until convergence is achieved.

\begin{figure}[h!]
    \centering
    \includegraphics[
        width=0.5\textwidth,    % 宽度为文本宽度的一半
        height=0.3\textheight,  % 高度设置
        keepaspectratio,        % 保持宽高比
        angle=0,                % 旋转角度
        scale=0.8               % 缩放比例
        ]{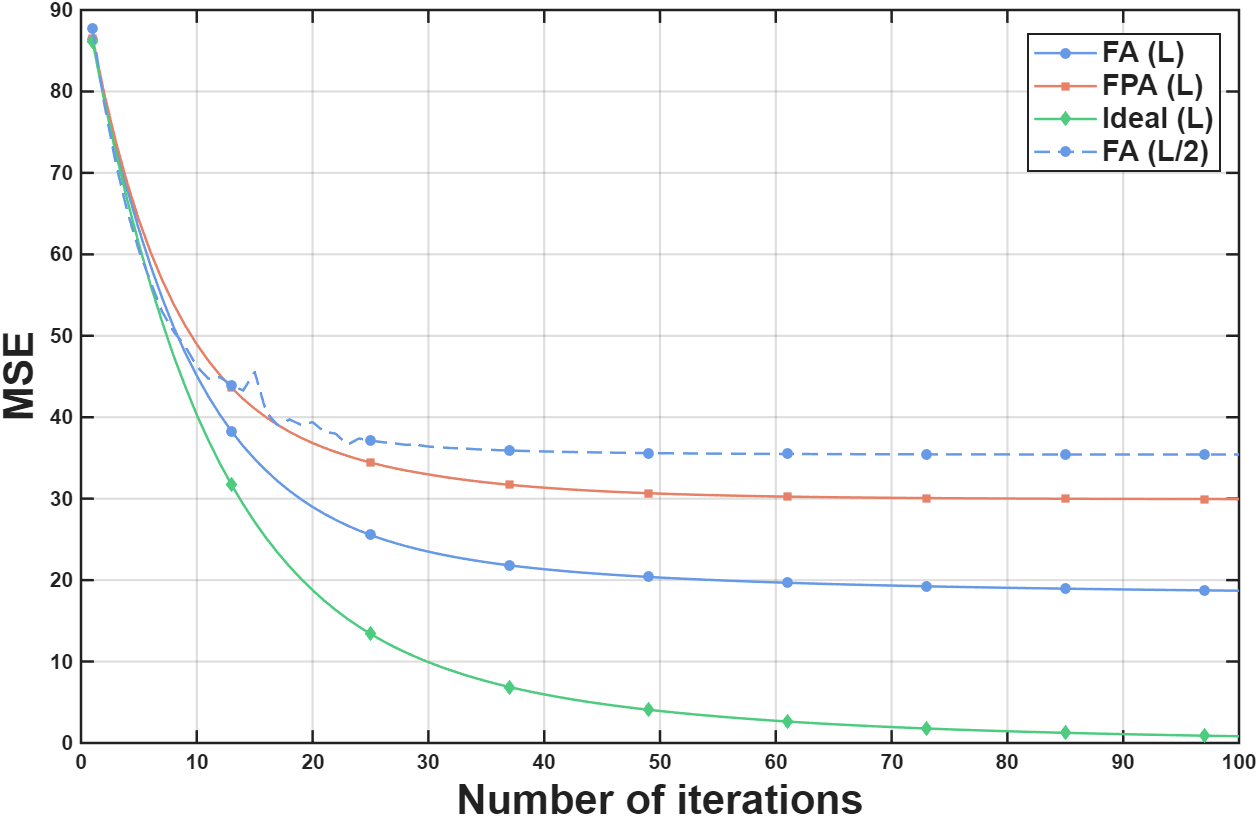}
    \caption{MSE versus the number of iterations, where $\textit{N} = 10$, and $\textit{K} = 100$.}
    \label{fig:placeholder}
\end{figure}

To show the convergence of Algorithm 3, we illustrate
the MSE as a function of the iterations in Fig. 2.  From
this figure, we observe that Algorithm 3 converges after a
few dozens of iterations. As can be seen from the figure, when the antenna movement range is reduced by half, the performance optimization drops significantly; and the proposed algorithm outperforms the FPA scheme.

Fig.3a depicts the MSE performance of the aforementioned scheme under varying numbers of FAs ($\textit{N}$). As observed,  the MSE under ideal hardware conditions achieves significantly lower values than the proposed scheme across all antenna configurations; the proposed design maintains a distinct advantage over the FPA benchmark; and all schemes exhibit progressively improved MSE performance (i.e., decreasing values) with increasing FA counts, consistent with theoretical expectations since deploying more FAs enhances spatial diversity and signal aggregation in AirComp.

\begin{figure}[h!]
    \centering
    \begin{minipage}[b]{0.24\textwidth}
        \centering
        \includegraphics[width=\textwidth]{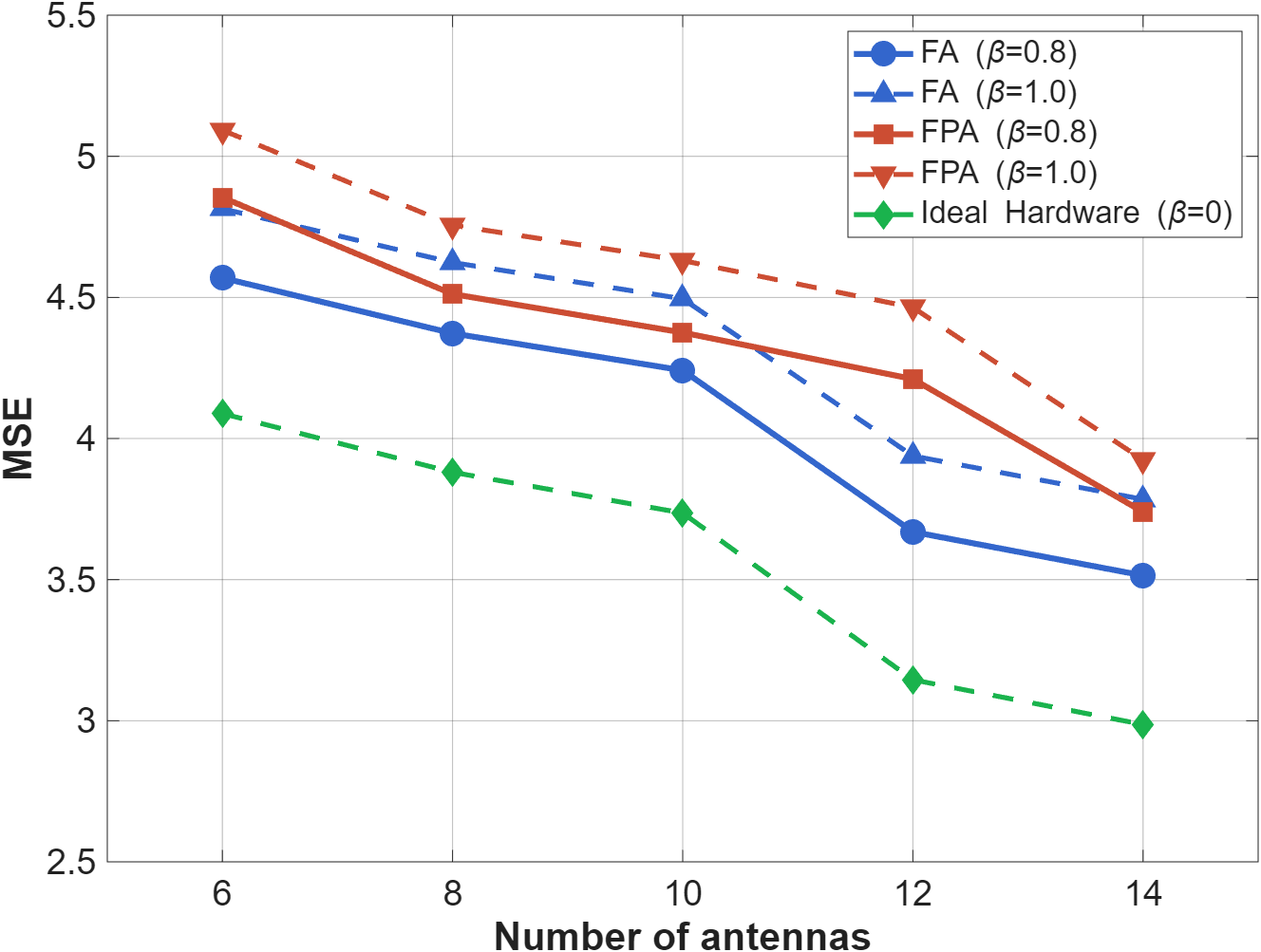}
        \par\vspace{2pt} % 添加一点垂直间距
        \centering (a)
        \label{a:a}
    \end{minipage}
    \hfill
    \begin{minipage}[b]{0.24\textwidth}
        \centering
        \includegraphics[width=\textwidth]{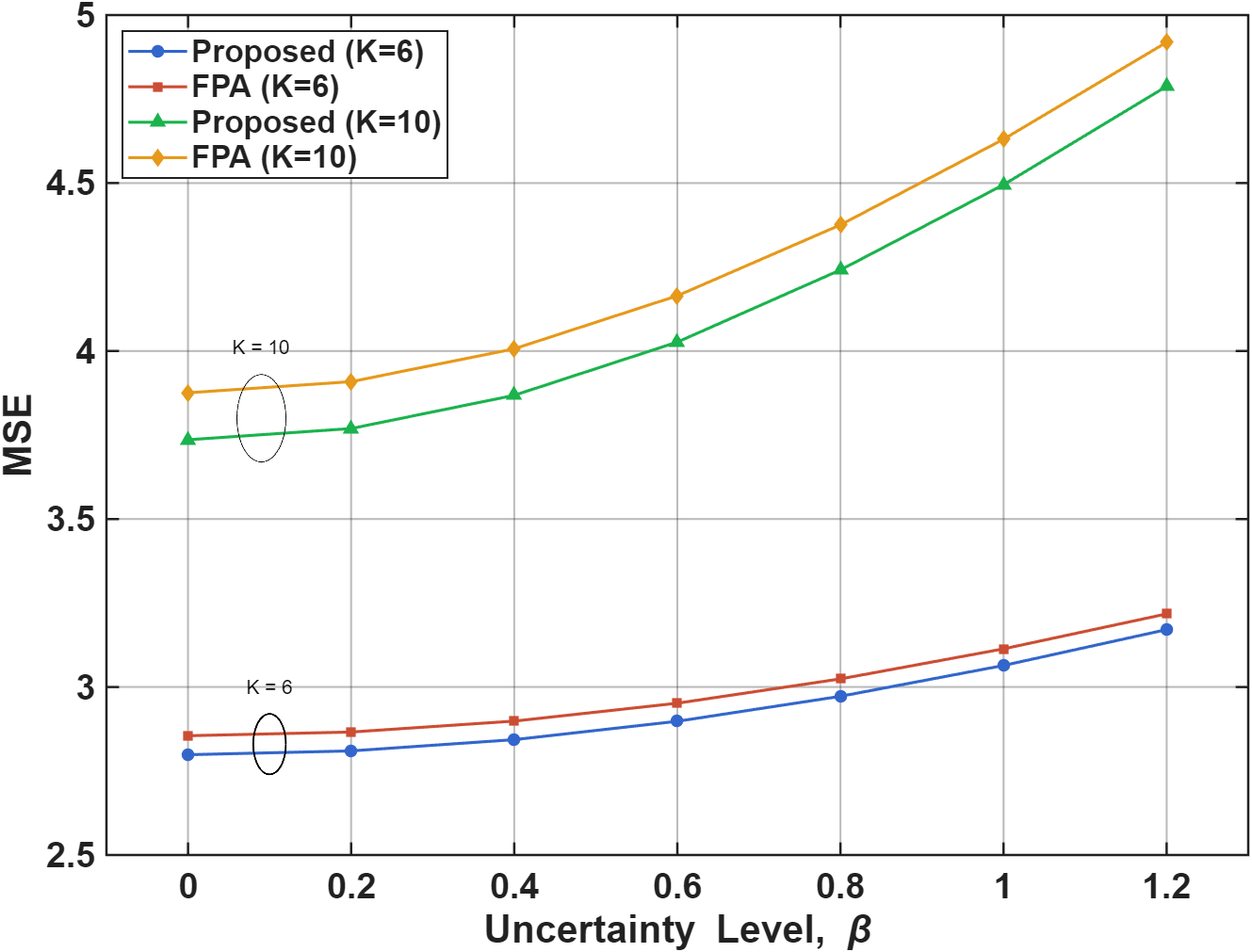}
        \par\vspace{2pt} % 添加一点垂直间距
        \centering (b)
        \label{fig:image2}
    \end{minipage}
    \caption{(a) MSE versus the number of FAs $\textit{N}$, where $\textit{K} = 10$. (b) MSE versus the uncertainty level under different numbers of users, where $\textit{N} = 10$.}
    \label{fig:both}
\end{figure}

Fig.3b compares the MSE performance between the proposed scheme and the FPA benchmark under different uncertainty levels ($\beta$) for user configurations $\textit{K} = 6$ and $\textit{K} = 10$. The proposed scheme consistently outperforms the FPA approach in both user scenarios, with the performance gap being more pronounced at $\textit{K} = 10$. As $\beta$ increases, all schemes exhibit degraded MSE performance, though the proposed method demonstrates superior robustness with a slower degradation rate. Additionally, systems with fewer users ($\textit{K} = 6$) achieve lower MSE than those with more users ($\textit{K} = 10$), which aligns with theoretical expectations. These results validate the effectiveness of the proposed design in mitigating uncertainty impacts while maintaining reliable performance.
\section{Conclusion}
This paper addresses hardware impairment scenarios by proposing an AirComp system based on FA arrays, with joint optimization of transceiver design and APV to minimize the MSE. The system modeling incorporates practical hardware-induced distortion noise and FA movement energy consumption constraints, formulated as an optimization problem.  Numerical results demonstrate that, compared to conventional FPA arrays, the proposed joint transceiver and FA position optimization significantly enhances the estimation accuracy and robustness of AirComp systems in communication environments with practical hardware impairments, providing a viable solution for future integrated communication and computation system design.


\begin{thebibliography}{1}
\bibitem{ref1}
Z. Yang et al. "On Privacy, Security, and Trustworthiness in Distributed Wireless Large AI Models (WLAM)." arXiv preprint arXiv:2412.02538 (2024).  
\bibitem{ref2}
W. Xu et al. "Edge Learning for B5G Networks With Distributed Signal Processing: Semantic Communication, Edge Computing, and Wireless Sensing." IEEE J. Sel. Topics Signal Process. 17.1 (2023): 9-39.
\bibitem{ref3}
G. Zhu et al., “Over-the-Air computing for wireless data aggregation in
massive IoT,” IEEE Wireless Commun, vol. 28, no. 4, pp. 57–65, Aug.
2021.
\bibitem{ref4}
Z. Wang et al. "Over-the-air computation for 6G: Foundations, technologies, and applications." IEEE Internet of Things Journal 11.14 (2024): 24634-24658.
\bibitem{ref5}
L. Qiao et al. "Massive digital over-the-air computation for communication-efficient federated edge learning." IEEE Journal on Selected Areas in Communications 42.11 (2024): 3078-3094.
\bibitem{ref6}
X. Jiao et al. "Task-oriented over-the-air computation for edge-device co-inference with balanced classification accuracy." IEEE Transactions on Vehicular Technology 73.11 (2024): 17818-17823.
\bibitem{ref7}
M. Fu et al. "UAV aided over-the-air computation." IEEE Transactions on Wireless Communications 21.7 (2021): 4909-4924.
\bibitem{ref8}
G. Zhu et al. "Over-the-air computing for wireless data aggregation in massive IoT." IEEE Wireless Communications 28.4 (2021): 57-65.
\bibitem{ref9}
K.-K. Wong and et al., “Fluid antenna systems,” IEEE Trans. Wireless Commun., vol. 20, no. 3, pp. 1950–1962, Nov. 2021.
\bibitem{ref10}
L. Zhu et al. "Movable antennas for wireless communication: Opportunities and challenges." IEEE Communications Magazine 62.6 (2023): 114-120.
\bibitem{ref11}
W. Ma et al. "Multi-beam forming with movable-antenna array." IEEE Communications Letters 28.3 (2024): 697-701.
\bibitem{ref12}
D. Zhang et al. "Fluid antenna array enhanced over-the-air computation." IEEE Wireless Communications Letters 13.6 (2024): 1541-1545.
\bibitem{ref13}
Pakravan, Saeid, et al. "Robust resource allocation for over-the-air computation networks with fluid antenna array." 2024 IEEE Globecom Workshops (GC Wkshps). IEEE, 2024.
\end{thebibliography}
\end{document}